\documentstyle[12pt]{article}
\newcommand{\beq}{\begin{equation}}
\newcommand{\eeq}{\end{equation}}
\newcommand{\beqa}{\begin{eqnarray}}
\newcommand{\eeqa}{\end{eqnarray}}
\newcommand{\ba}{\begin{array}}
\newcommand{\ea}{\end{array}}
\newcommand{\CR}{\nonumber \\}
\newcommand{\no}{\nonumber \\}
\newcommand{\pa}{\partial}
\newcommand{\A}{\alpha}
\newcommand{\B}{\beta}
\newcommand{\D}{\delta}

\newcommand{\La}{\Lambda}

\newcommand{\lm}{\lambda}

\newcommand{\th}{\vartheta}
\newcommand{\bra}{\langle}
\newcommand{\ket}{\rangle}
\newcommand{\half}{{1 \over 2}}
\renewcommand{\thefootnote}{\fnsymbol{footnote}}
\newcommand{\Lf}{\Lambda_{N_f}}

\begin{document}
\begin{titlepage}
\begin{flushright}
{\tt hep-th/9806015} \\
UTHEP-382 \\
June, 1998
\end{flushright}
\vspace{0.5cm}
\begin{center}
{\Large \bf 
Donaldson-Witten Functions \\ 
of Massless N=2 Supersymmetric QCD
\par}
\lineskip .75em
\vskip2.5cm
{\large Hiroaki Kanno}
\vskip 1.5em
{\large\it Department of Mathematics, Faculty of Science \\
Hiroshima University \\
Higashi-Hiroshima 739-8526, Japan}
\vskip1cm
{\large Sung-Kil Yang}
\vskip 1.5em
{\large\it Institute of Physics, University of Tsukuba \\
Ibaraki 305-8571, Japan}
\end{center}
\vskip2cm
\begin{abstract}
We study the Donaldson-Witten function in four-dimensional topological gauge 
theory which is constructed from $N=2$ supersymmetric $SU(2)$ gauge theory 
with $N_f < 4$ massless fundamental hypermultiplets. 
When $N_f = 2,3$, the strong-coupling singularities with multiple massless
monopoles appear in the moduli space (the $u$-plane) of the Coulomb branch. 
We show that the invariants made out of such singularities exhibit a 
property which is similar to the one expected for four-manifolds of
generalized simple type.
\end{abstract}
\end{titlepage}
\baselineskip=0.7cm

\renewcommand{\thefootnote}{\arabic{footnote}}
\setcounter{footnote}{0}

\section{Introduction}

Recent rapid developments in non-perturbative analysis of four-dimensional 
$N=2$ supersymmetric gauge theory \cite{SeWi1}\cite{SeWi2} have
deepened our understanding of Donaldson theory which is known to be formulated
as a twisted version of $N=2$ supersymmetric Yang-Mills theory \cite{W1}.
In the strong-coupling approach to Donaldson theory the non-abelian problem
is replaced with the abelian problem which is much more tractable \cite{W2}.
As a result new insights into four-manifolds have been 
obtained \cite{Don}\cite{Mor}.

In a recent important paper Moore and Witten investigated Donaldson invariants
for manifolds with $b_2^+=1$ \cite{MoWi}. To this aim the contribution from the
entire Coulomb branch ({\it i.e.} the $u$-plane integral) of $N=2$ Yang-Mills 
theory has been analyzed in great detail. It is shown clearly how
elliptic modular functions enter Donaldson theory. 
The Donaldson-Witten function is a generating function of Donaldson
invariants and in the case $b_2^+ >1$ all contributions come from
the strong-coupling singularities. It is interesting that considering 
the wall-crossing phenomenon for $b_2^+=1$ at strong-coupling singularities
enables us to calculate the Donaldson-Witten function in the case $b_2^+ >1$. 
The work has then been 
generalized in several directions \cite{MM1}\cite{MM2}\cite{LNS}\cite{MM3}.
There is also a remarkable connection of topological gauge theory 
to integrable systems \cite{GMMM}\cite{MM2}\cite{Tak}\cite{EMM}.

In this paper we study the Donaldson-Witten function when a dual abelian gauge
field couples to $k$ massless monopole fields with $k \geq 1$ at the
strong-coupling singularity. This situation arises in the $N=2$ $SU(2)$
theory with massless $N_f$ hypermultiplets in the $SU(2)$ doublet \cite{SeWi2}.
The moduli space possesses singularities where $k=2^{N_f-1}$ massless
monopoles appear. Our result provides another generalization of the relation
between invariants of four-manifolds and Seiberg-Witten theory.

This paper is organized as follows. In section two, we review 
Seiberg-Witten theory of massless $N=2$ $SU(2)$ QCD
and relevant mathematics of the Seiberg-Witten curve.
We evaluate in section three the Seiberg-Witten contributions
to the Donaldson-Witten function. The Seiberg-Witten contribution from 
the strong-coupling singularities with a single massless monopole 
has been obtained in \cite{MoWi}. We complete the analysis in \cite{MoWi}
by generalizing it to the singularity with $k>1$ massless monopoles
(or dyons) that appears when $N_f=2,3$. We also show that even if
we have multiple massless monopoles, there are only finitely many
$Spin^c$-structures which have non-vanishing contributions.
In section four we apply our formula to the simplest example
of a $K3$ surface. For $N_f=2,3$, we observe contributions from 
higher order terms in a series expansion near singularities.
These contributions make the Donaldson-Witten function more
complicated than the case of $N_f = 0, 1$ where only a single
massless monopole appears. We note that they are similar to
what we expect for a hypothetical four-manifold that is not
of simple type. The final section is devoted to discussions.
Many useful formulas of elliptic modular functions associated to
the Seiberg-Witten curve of massless $N=2$ $SU(2)$ QCD are collected
in Appendix.

\section{$N=2$ $SU(2)$ QCD and elliptic curves}

\renewcommand{\theequation}{2.\arabic{equation}}\setcounter{equation}{0}

Let us first collect relevant low-energy properties of the $N=2$ $SU(2)$ 
theory with massless $N_f$ fundamental hypermultiplets. The Coulomb branch of
the theory is parametrized by $u$ which is a gauge invariant expectation
value of the $SU(2)$ adjoint Higgs field. The symmetry 
${\bf Z}_{4-N_f}$ acts on
the complex $u$-plane when $N_f \geq 1$. At a generic point on the $u$-plane
there exists an $N=2$ $U(1)$ vector multiplet. At singularities there appear
extra massless monopole (or dyon) hypermultiplets. They belong to the spinor
representation of the $SO(2N_f)$ global flavor symmetry. Thus the 
multiplicity of massless monopoles equals $k=2^{N_f-1}$. Note that for 
$N_f=3$ no symmetry acts on the $u$-plane and there is a singularity 
associated with a massless $SO(6)$ singlet dyon.

All the low-energy physics of the Coulomb branch is encoded
in the geometry of elliptic curves. According to \cite{SeWi2} the relevant
curves are given by
\beqa
&& y^2=x^2(x-u)+{1\over 4}\La_0^4x, \hskip10mm N_f=0, \CR
&& y^2=x^2(x-u)-{1\over 64}\Lf^{2(4-N_f)}(x-u)^{N_f-1},
\hskip10mm N_f=1,\, 2,\, 3,
\label{curve}
\eeqa
where $\Lf$ is the dynamical scale. The change of variables $y=4Y$ and
$x=4(X+{u\over 12}+{\La_3^2 \over 12\cdot 64}\delta_{N_f,3})$ renders the
curves into the Weierstrass form
\beq
Y^2=4X^3-g_2X-g_3=4(X-e_1)(X-e_2)(X-e_3).
\label{cubic}
\eeq
To deal with the period integrals on the curve it is helpful to
employ the uniformization map $(\wp(z),\wp'(z))=(X,Y)$ where $\wp(z)$
is the Weierstrass $\wp$-function \cite{AMZ}. The well-known relations are
$\wp(\omega_\nu)=e_\nu$, $\wp'(\omega_\nu)=0$ $(\nu =1,2,3)$ where
$\omega_\nu$ are the half periods obeying $\omega_1+\omega_2+\omega_3=0$.
The periods $(2\omega_1, 2\omega_3)$ of a torus are obtained as
\beqa
&& 2\omega_1=\oint_A {dX\over Y}=\int_{e_3}^{e_2} 
{dX \over \sqrt{(X-e_1)(X-e_2)(X-e_3)}}, \no
&& 2\omega_3=\oint_B {dX\over Y}=\int_{e_1}^{e_2}
{dX \over \sqrt{(X-e_1)(X-e_2)(X-e_3)}}.
\eeqa
The modular ratio of the torus $\tau =\omega_3/\omega_1$
determines the effective gauge coupling constant $\tau =\theta_{eff}/\pi+
8\pi i/g^2_{eff}$ \cite{SeWi2}.

The roots of the cubic (\ref{cubic}) are expressed in terms of the Jacobi theta
functions
\beq
e_1=\left( {\pi \over 2\omega_1} \right)^2 {1\over 3}(\th_3^4+\th_4^4),
\hskip5mm 
e_2=\left( {\pi \over 2\omega_1} \right)^2 {1\over 3}(\th_2^4-\th_4^4), 
\hskip5mm
 e_3=-\left( {\pi \over 2\omega_1} \right)^2 {1\over 3}(\th_2^4+\th_3^4),
\label{eroots}
\eeq
where
\beq
\th_2=\sum_{n \in {\bf Z}}q^{\half \left(n+\half \right)^2}, \hskip10mm
\th_3=\sum_{n \in {\bf Z}}q^{\half n^2}, \hskip10mm
\th_4=\sum_{n \in {\bf Z}}(-1)^n q^{\half n^2}
\eeq
with $q=e^{2\pi i\tau}$. From (\ref{cubic}) and (\ref{eroots}) one has
\beq
g_2={2\over 3} \left( {\pi \over 2\omega_1} \right)^4 f, \hskip10mm
g_3={4\over 27} \left( {\pi \over 2\omega_1} \right)^6 h,
\label{g2g3}
\eeq
where
\beqa
&& f=\th_2^8+\th_3^8+\th_4^8 =2(1+240q+2160q^2+ \cdots ), \no
&& h=(\th_3^4+\th_4^4)(\th_4^4-\th_2^4)(\th_2^4+\th_3^4)
=2(1-504q-16632q^2+\cdots ). 
\label{fandh}
\eeqa

The discriminant $\Delta$ of the curve (\ref{cubic}) is given by
\beq
\Delta =g_2^3-27g_3^2=2^{12} \left( {\pi \over 2\omega_1} \right)^{12}
\eta^{24}(\tau),
\label{discri}
\eeq
where $\eta (\tau)=q^{1/24}\prod_{n=1}^\infty (1-q^n)$.
Explicitly we have 
\beq
\Delta (u)=\left\{ \begin{array}{ll}
\displaystyle{\hskip1.5mm {\La_0^8\over 2^{12}} (u-\La_{0}^2)(u+\La_{0}^2)}, 
& \qquad N_{f}=0, \\
\displaystyle{-{\La_1^6\over 2^{12}} \Bigl(u^3+27 \La_1^6/256 \Bigr)}, 
& \qquad N_{f}=1, \\
\displaystyle{\hskip1.5mm {\La_2^4\over 2^{12}}\Bigl(u-\La_2^2/8\Bigr)^2
\Bigl(u+\La_2^2/8 \Bigr)^2}, 
& \qquad N_{f}=2, \\
\displaystyle{-{\La_3^2 \over 2^{12}} u^4\Bigl(u-\La_3^2/256\Bigr)}, 
& \qquad N_{f}=3.
\end{array}
\right.
\eeq

The Seiberg-Witten differential for the $SU(2)$ theory with massless
fundamental matters takes the form \cite{IY}
\beq
\lambda_{SW}={\sqrt{2}\over 8\pi}{2u-(4-N_f)x \over y}dx.
\eeq
The period integrals are evaluated to be \cite{AMZ}
\beqa
&& a(u)=\oint_A \lambda_{SW}={\sqrt{2} \over \pi} \left[ \left(
(2+N_f)u-{\La_3^2 \over 64}\delta_{N_f,3} \right) {\omega_1 \over 12}
+(4-N_f)\zeta (\omega_1) \right], \no
&& a_D(u)=\oint_B \lambda_{SW}={\sqrt{2} \over \pi} \left[ \left(
(2+N_f)u-{\La_3^2 \over 64}\delta_{N_f,3} \right) {\omega_3 \over 12}
+(4-N_f)\zeta (\omega_3) \right],
\label{period}
\eeqa
where $\zeta'(z)=-\wp(z)$. Taking the derivative with respect to $u$, 
we get\footnote{We note in passing that combining (\ref{discri}) and
(\ref{holo}) yields ${da \over du}={\sqrt{2} \over 4} \eta^2 \Delta^{-1/12}$.
This expression is useful in the $F$-theory consideration \cite{Sen}.}
\beq
{da \over du}={\sqrt{2} \over 8} {2\omega_1 \over \pi}, \hskip10mm\
{da_D \over du}={\sqrt{2} \over 8} {2\omega_3 \over \pi}.
\label{holo}
\eeq

It is known that the periods $\Pi =(a(u), a_D(u))$ obey the Picard-Fuchs 
equation \cite{KLT}\cite{IY}
\beq
P_{N_f}(u) {d^2 \Pi \over du^2}+\Pi =0,
\label{pf}
\eeq
where
\beq
P_{N_f}(u)=\left\{ \begin{array}{ll}
\displaystyle{4 (u^2-\La_{0}^4)}, & \qquad N_{f}=0, \\
\displaystyle{4u^2+{27 \La_1^6\over 64u}}, & \qquad N_{f}=1, \\
\displaystyle{\ 4\Big(u^2-\La_2^4/ 64 \Big)}, & \qquad N_{f}=2, \\
\displaystyle{4u\Big(u-\La_3^2/256 \Big)}, & \qquad N_{f}=3.
\end{array}
\right.
\label{pfcoef}
\eeq
The absence of the first derivative term in (\ref{pf}) implies that the 
Wronskian for the solutions is constant. A useful identity then comes 
out \cite{MSEY}
\beq
a {da_D \over du}-a_D {da\over du}=i{4-N_f \over 4\pi}.
\label{wronski}
\eeq
If we use the Legendre's relation 
$\zeta (\omega_1)\, \omega_3-\zeta (\omega_3)\, \omega_1= \pi i /2$,
this is an immediate consequence from (\ref{period}) and (\ref{holo}).
It is now easy to verify from (\ref{pf}) and (\ref{wronski}) that
\beq
{d\tau \over du}=i {8(4-N_f) \over \pi P_{N_f}(u)} 
\left({\pi \over 2\omega_1}\right)^2.
\label{jacobi}
\eeq
where we have used the basic relation $\tau =da_D/da$.

\section{Donaldson-Witten functions}

\renewcommand{\theequation}{3.\arabic{equation}}\setcounter{equation}{0}

Let us now turn to topological gauge theory on a four-manifold $X$.
The Seiberg-Witten theory provides a quantum field theoretical approach to
the computation of Donaldson invariants of $X$. In the following we
assume that $X$ is simply connected, or $b_1=b_3=0$, where $b_k$ denotes
the $k$-th Betti number. A harmonic two form on $X$ can be decomposed
into a sum of self-dual and anti-self-dual components. We have
$b_2 = b_2^+ + b_2^-$ with $b_2^+$ and $b_2^-$ being the dimensions
of the spaces of self-dual harmonic two forms and anti-self-dual two
forms, respectively. The Euler characteristic of $X$ is $\chi = 2 + b_2^+ +
b_2^-$ and the signature is $\sigma = b_2^+ - b_2^-$. 
The second homology and cohomology group $H_2 (X, {\bf Z})$ and $H^2 (X,
{\bf Z})$ 
have a ring structure by the intersection form.  This ring structure
is the classical topological invariants for the classification of
four-manifolds. The Donaldson invariants produce a powerful tool beyond the
classical cohomology ring.

The Donaldson-Witten function $Z_{DW}$ of a four-manifold $X$ 
is a generating function of Donaldson invariants. 
In the framework of topological gauge theory
it is given as a generating function of correlation functions \cite{W1},
\beq
Z_{DW} (p, S) = \left\bra \exp \Bigl( p {\cal O} + I(S) \Bigr)\right\ket~, 
\eeq
where ${\cal O}$ and $I(S)$ are certain operators (observables) associated
with a point $p \in H_0 (X, {\bf Z})$ and a surface $S \in H_2(X, {\bf Z})$.
The vacuum expectation value of ${\cal O}$ is identified with
the quantum moduli parameter $2u = \bra {\cal O} \ket$ in Seiberg-Witten
theory. The two form observable $I(S)$ in the effective $U(1)$ topological
theory creates the contact term interaction $T(u)$ at the intersection
points of two surfaces $S_1$ and $S_2$ \cite{MoWi}.
$Z_{DW}$ is a sum of the $u$-plane integral and the contribution from 
the Seiberg-Witten invariants;
\beq
Z_{DW} = Z_u + Z_{SW}~.
\label{dwinv}
\eeq
We consider a topological theory which is a
twisted version of $N=2$ $SU(2)$ gauge theory with massless $N_f$ fundamental
matters. For this theory the $u$-plane integral $Z_u$ takes the form
which is obtained by setting the bare quark masses equal to zero in the
expression given in \cite{MoWi}. 
On the other hand the SW contribution $Z_{SW}$ differs substantially,
since the massless monopoles appear with multiplicity $k=2^{N_f-1}$
at strong-coupling singularities in the massless $N=2$ $SU(2)$ QCD.
Thus our main concern henceforth is to fix $Z_{SW}$.

In (\ref{dwinv}), when $b_2^+ > 1$, there is no contribution from the $u$-plane
integral $Z_u$ \cite{MoWi}. 
The $u$-plane integral, however, is responsible for the peculiar phenomena 
we encounter in the case of $b_2^+ =1$ such
as the wall crossing. According to Moore and Witten \cite{MoWi},
one can derive the relation of Donaldson and Seiberg-Witten
invariants from the wall crossing at strong-coupling 
singularities. For each class $\lm \in H_2(X, {\bf Z}) + \half w_2(E)$,
the wall crossing formula of $Z_u$ at a zero $u = u_*$ of 
the discriminant $\Delta(u)$ of the Seiberg-Witten curve is
\beqa
Z_{u,+} - Z_{u,-} &=& 2\sqrt{2}~e^{2\pi i \lm_0^2}~
 (-1)^{(\lm -\lm_0)\cdot w_2(X)} \CR
& & \A^\chi \B^\sigma \left[ q^{-\lm^2/2} \Big( \frac{du}{d\tau} \Big)
\Big( \frac{da}{du} \Big)^{1-\chi/2} \Delta^{\sigma/8}
\exp \Bigl( 2pu + S^2 T(u) -i \frac{du}{da}(S, \lm) \Bigr)
\right]_{q^0}, \no
&&
\eeqa
where $\lm_0 \in H_2(X, {\bf Z}) + \half w_2(E)$ has been introduced
to define an orientation of the instanton moduli space. (For $SU(2)$ theory,
$w_2(E)=0$ and one may take a canonical choice $\lm_0 = 0$.)
If $X$ is a spin manifold, that is $w_2(X)=0$, then the sign factor
disappears.  Here the factors $\alpha^\chi \beta^\sigma (du/da)^{\chi /2}
\Delta^{\sigma /8}$ have appeared as the measure factor for the $u$-plane 
integral \cite{W3}\cite{MoWi}. Near the zero $u=u_*$ with multiplicity $k$
we have introduced a good local coordinate $q$ by
\beq
(u - u_*)^k = \kappa_* q + O(q^2)~,
\eeq
and the subscript $q^0$ means taking the coefficient of the $q^0$ term
in the $q$ expansion. Hence, if the leading power of the $q$ expansion
in $q \rightarrow 0$ is positive, then there is no wall crossing 
contribution from $\lm$. 

Let us check when $\lm$ gives non-trivial 
contribution. From (\ref{holo}) and (\ref{jacobi}) we have
\beq
\frac{du}{d\tau}  = 
\frac{4\pi}{i(4-N_f)} \Big( \frac{da}{du} \Big)^2 P_{N_f}(u)~.
\eeq
Since $P_{N_f}(u)$ has a simple zero at $u=u_*$ (see (\ref{pfcoef})),
$P_{N_f}(u) \sim \Delta(u)^{1/k}$ near the singularity. 
Using the following expansions of the period and the
discriminant,
\beqa
 \frac{da}{du}  &=& \Big( \frac{da}{du} \Big)_* + (u-u_*) 
\Big( \frac{d^2a}{du^2} \Big)_* + \cdots~, \CR
\Delta(u) &=& \Delta_*^{(k)}(u-u_*)^k + \cdots~,
\eeqa
we see the leading term in the $q$ expansion is 
\beq
q^{-\lm^2/2} \cdot q^{1/k} \cdot q^{\sigma/8} + \cdots~, \quad (q
\rightarrow 0)~.
\label{qexp}
\eeq
Here and henceforth $a$ should be a good local coordinate at strong-coupling 
singularity. It is a linear combination of $a$ and the dual
variable $a_D$ in general. (See a discussion in section 4 for detail.)
For example, at the monopole singularity $a$ is understood as $a_D$.
For a non-vanishing contribution the power in (\ref{qexp})
should be non-positive, that is,
\beq
k \lambda^2 -2 - \frac{k\sigma}{4} \geq 0~.
\label{bound}
\eeq

\subsection{Measure factors}

Near the singularity, the dual theory is described as an effective $U(1)$
gauge theory. There are $k$ massless $U(1)$ hypermultiplets for 
the $k$-th order zero of $\Delta$.
The twisted action is topological up to BRST exact terms 
and it would have the form
\beq
L = \{ Q_B, W \} + \int_X \left( c(u) F \wedge F + p(u) {\rm Tr}~R \wedge R
+  \ell(u) {\rm Tr}~R \wedge \widetilde R \right)~.
\eeq
Note that the overall couplings of the topological terms may depend on the
vacuum moduli $u$. Among the topological terms the $F \wedge F$ term 
is the descendant of the prepotential ${\cal F}_M$ of
the dual photon and monopoles (dyons). The coefficient $c(u)$ is related
to the effective coupling $\tau_M$, the second derivative of ${\cal F}_M$.
The other terms with the curvature $R$ are the background gravitational
effect due to the integration over massive fermions. In the path integral
these couplings induce the following measure factor,
\beq
C(u)^{\lm^2/2}~P(u)^{\sigma/8}~L(u)^{\chi/4}~.
\eeq
Following the approach of \cite{MoWi}, we will determine the measure factor
by matching the wall crossing formula of $Z_u$ and that of $Z_{SW}$. 
This means that there is no jump in the invariants at 
any strong coupling singularity;
\beq
\delta Z_{DW} = \delta Z_u + \delta Z_{SW} = 0~.
\eeq
This matching is imposed only when $b_2^+ =1$. But, since
the measure factors are universal in the sense that they do not depend on
the manifold on which topological theory is defined, we can use the same
measure factors also for the case $b_2^+ >1$. 

If we assume that the jump of the SW invariant $SW(\lm)$ at the wall is 
$\pm k$,\footnote{Note that there are $k$ massless hypermultiplets.}
the wall crossing of the Seiberg-Witten invariants is
\beqa
\Delta \bra e^{p {\cal O} + I(S)} \ket &=&
\pm {\rm Res}_{a = a_*}~\biggl[ 2k~e^{2\pi i(\lm_0\cdot \lm + \lm_0^2)}
\frac{da}{(a - a_*)^{1 + d_\lm/2}} \CR
& & C(u)^{\lm^2/2}~P(u)^{\sigma/8}~L(u)^{\chi/4}~\exp \Bigl(
2pu + S^2 T(u) -i \frac{du}{da}(S, \lm) \Bigr)
\biggr], \label{SWWC}
\eeqa
where $\lm =c_1 (L^2)/2$, $\D =(\chi + \sigma)/4$ and
\beq
d_\lm = k \left(\lm^2 - \frac{\sigma}{4}\right) - 2 \D
\label{index}
\eeq
is the formal dimension of the moduli space of a generalized
Seiberg-Witten monopole equation with $k$ massless hypermultiplets;
see (\ref{kmono1}) and (\ref{kmono2}) for explicit forms.
When $b_2^+ = 1$, or $\chi + \sigma =4$, the condition (\ref{bound}) 
for non-trivial wall crossing of $Z_u$ agrees with the condition that
the dimension of the moduli space is non-negative $d_\lm \geq  0$. 
That is the condition under which we have non-trivial residue in (\ref{SWWC}). 
Thus we can consistently 
require the matching of two wall crossing formulae, which implies
\beqa
&& \hskip5mm C(u)^{\lm^2/2}~P(u)^{\sigma/8}~L(u)^{1- \sigma/4} \no
&& =\sqrt{2} k^{-1} \A^{4 - \sigma} \B^\sigma (a- a_*)^{\frac{k}{2}(\lm^2 - 
\frac{\sigma}{4})} \frac{d (\log q)}{da} q^{-\lm^2/2}
\Big( \frac{du}{d\tau} \Big)  \Big( \frac{da}{du} \Big)^{\sigma/2 -1} 
\Delta^{\sigma/8} \CR
&& = 2\sqrt{2} i\pi k^{-1} \A^{4 - \sigma} \B^\sigma 
(a- a_*)^{\frac{k}{2}(\lm^2 - \frac{\sigma}{4})}
q^{-\lm^2/2} \Big( \frac{da}{du} \Big)^{\sigma/2 -2} 
\Delta^{\sigma/8}~.
\eeqa
We obtain the following results for the measure factors,
\beqa
&& C(u) = \frac{1}{q} (a - a_*)^k, \no
&& P(u) = - 8 \pi^2 k^{-2} \B^8 (a - a_*)^{-k} \Delta(u), \no
&& L(u) = 2\sqrt{2} i\pi k^{-1} \A^4 \Big( \frac{du}{da} \Big)^2. 
\eeqa

These measure factors derived from the condition that there is
no discontinuity of $Z_{DW}$ at a strong-coupling singularity might be 
deduced from the following argument.\footnote{It is better to have
some qualitative reasoning which is independent of 
a detail of the wall crossing formula.} First we note that the logarithm of 
the measure factor $C(u)$ gives the effective coupling;
\beq
\tau_M = \tau_D - \frac{k}{2\pi i} \log (a - a_*)~,
\label{eff}
\eeq
which is indeed the desired relation we expect from the existence of 
$k$ massless matters.
Up to numerical constants the only difference of the gravitational measure
factors from those for the $u$-plane integral is the factor $(a- a_*)^{-k}$
in $P(u)$. This also has an explanation in view of the existence 
of $k$ hypermultiplets, which is the additional massless excitations 
compared to the topological theory
at a generic point on the $u$-plane. Recall that the gravitational
measure factor for effective topological theory is fixed
by counting the $R$ charge \cite{W3}. The contribution of the
hypermultiplet to the $R$ charge is twice the index of
the Dirac operator
\beq
2 \cdot ({\rm index}~D) =   \lambda^2 - \frac{\sigma}{4} ~.
\eeq
Hence the gravitational part of $k$ hypermultiplets is $- (k/4)\sigma$.
We see that there is no change in $L(u)$, but $P(u)$ should have 
an additional factor with the $R$ charge $- 2k$. 
Now the factor $(a - a_*)^{-k}$ does the job, since $(a - a_*)$ carries the
$R$ charge $+2$.

\subsection{Seiberg-Witten contributions}

Using these measure factors, we obtain the following universal 
form of the Seiberg-Witten contribution to the invariants of
four-manifolds;
\beqa
Z_{SW} &=& \sum_\lm SW(\lm) \,
{\rm Res}_{a = a_*}~\biggl[ 2~e^{2\pi i(\lm_0\cdot \lm + \lm_0^2)}
\frac{da}{(a - a_*)^{1 + d_\lm/2}} \CR
& & \Bigl( \frac{(a-a_*)^k}{q}\Bigr)^{\lm^2/2}\ 
(-8 \pi^2 k^{-2} \B^8)^{\sigma/8}\ \Bigl( \frac{\Delta^{\sigma/8}}
{(a - a_*)^{k\sigma/8}}\Bigr) \ (2\sqrt{2} \pi i k^{-1} \A^4)^{\chi/4}\ 
\Bigl(\frac{du}{da}\Bigr)^{\chi/2} \CR
& & \exp \Bigl(
2pu + S^2 T(u) -i \frac{du}{da}(S, \lm) \Bigr)
\biggr]  \CR
&=& \sum_\lm SW(\lm)\, 
{\rm Res}_{a = a_*}~\biggl[ 2~e^{2\pi i(\lm_0\cdot \lm + \lm_0^2)}
\frac{da}{(a - a_*)^{1-\delta}} (2\sqrt{2} \pi i k^{-1})^\delta \A^\chi
\B^\sigma \CR
& & q^{-\lm^2/2} \Delta^{\sigma/8} \Bigl(\frac{du}{da}\Bigr)^{\chi/2} 
\exp \Bigl(
2pu + S^2 T(u) -i \frac{du}{da}(S, \lm) \Bigr) \biggr]~.
\eeqa
Here $SW(\lm)$ stands for the Seiberg-Witten invariant. Denoting as
${\cal M}_\lm$ the moduli space of solutions to the generalized 
monopole equations (\ref{kmono1}) and (\ref{kmono2})
with given $\lm$, we have
\beq
SW(\lm) = \bra \widetilde{a}^n \ket_\lm = 
\int_{{\cal M}_\lm} \widetilde{a}^n
\label{SWinv}
\eeq
for $d_\lm =2n$. In (\ref{SWinv})
we have used $\widetilde{a}$ to emphasize that it is a linear
combination of $a$ and $a_D$ according to the charge of 
massless excitation at the singularity.
Substituting the following relation derived from the leading part 
of the $q$-expansion,
\beqa
(u - u_*) &=& \kappa_* q + O(q^2)~, \CR
(a- a_*) &=&  (\frac{da}{du})_* (u-u_*) + \cdots 
= \kappa^{1/k} (\frac{da}{du})_* q^{1/k} + \cdots~, \CR
\frac{da}{a-a_*}  &=&  \frac{dq}{kq}~, \CR
\Delta &=& \Delta_*^{(k)}(u- u_*)^k + \cdots = \Delta_*^{(k)} \kappa_* q +
\cdots~, 
\eeqa
we have
\beqa
Z_{SW} 
&=& \sum_\lm SW(\lm)  2~e^{2\pi i(\lm_0\cdot \lm + \lm_0^2)}
(2\sqrt{2} \pi i k^{-1})^\delta \A^\chi \B^\sigma  \no
&& \biggl[ 
\frac{1}{kq^{d_\lm/2k}} (\kappa_*)^{\delta/k + \sigma/8} 
(\Delta_*^{(k)})^{\sigma/8}
\Bigl(\frac{da}{du}\Bigr)_*^{\delta-\chi/2} 
\exp \Bigl( 2pu_* + S^2 T_* -i \Bigl( \frac{du}{da}\Bigr)_* 
(S, \lm) \Bigr) \no
&&  + \hbox{sub-leading terms} \biggr]_{q^0}~.
\label{ours}
\eeqa
Note that when $d_\lm$ is positive, the sub-leading terms contribute to
$Z_{SW}$. This situation is analogous to the case of four-manifolds of
generalized simple type \cite{KM}\cite{MoWi}.

A four-manifold $X$ is called simple type (in the sense of Seiberg-Witten) if
$SW(\lm)=0$ for $d_\lm$ which is strictly positive. 
Let us assume that the discriminant has a simple zero at $u= u_*$. 
In this case $Z_{SW}$ for a manifold of simple type is much simplified to
\beqa
Z_{SW} &= & \sum_\lm SW(\lm)  2~e^{2\pi i(\lm_0\cdot \lm + \lm_0^2)}
(2\sqrt{2} \pi i)^\delta \A^\chi \B^\sigma \CR
& &  (\kappa_*)^{\delta + \sigma/8} 
(\Delta'_*)^{\sigma/8}
\Bigl(\frac{da}{du}\Bigr)_*^{\delta-\chi/2} 
\exp \Bigl( 2pu_* + S^2 T_* -i (\frac{du}{da})_* (S, \lm) \Bigr)~.
\label{simple}
\eeqa
Notice the relations derived from (\ref{g2g3}) and (\ref{discri})
\beq
\Bigl( \frac{da}{du} \Bigr)_*^2={1 \over 12^2}{g_2(u_*)\over g_3(u_*)},
\hskip10mm
\kappa_* \Delta'_* =2^{12} \Bigl( {\pi \over 2\omega_1}\Bigr)_*^{12}.
\eeq
Hence,
\beq
\Bigl( \kappa_* \Delta'_* \Bigr)^{\sigma/8}
\Bigl( \frac{da}{du} \Bigr)_*^{3\sigma/2} = 2^{-9\sigma/4}
\eeq
and using $\delta-\chi/2 = 3\sigma/2 - (\delta+\sigma)$, we have
\beq
(\kappa_*)^{\delta + \sigma/8} 
(\Delta'_*)^{\sigma/8}
\Bigl(\frac{da}{du}\Bigr)^{\delta-\chi/2} =
2^{-9 \sigma/4} (\kappa_*)^\delta 
\Bigl(\frac{da}{du}\Bigr)^{- (\delta+\sigma)}~.
\eeq
Thus it is seen that (\ref{simple}) for $k=1$ agrees with (11.28) of 
\cite{MoWi} and our formula (\ref{ours}) generalizes their expression.

\subsection{Vanishing theorem}

It is known that there are only finite number of isomorphism
classes of the line bundle $\lm$ such that the Seiberg-Witten
invariant $SW(\lm)$ is non-vanishing. This is a vanishing
theorem in \cite{W2}.  We can generalize the estimate implying the
vanishing theorem to the case of more than one massless
hypermultiplet.

If there are $k$ massless hypermultiplets whose bosonic
components are $M_\alpha~(\alpha = 1, 2, \ldots, k)$, 
the monopole equations are generalized to
\beqa
&&  F_{ij}^+ = -\frac{i}{2} \sum_{\alpha=1}^{k} {\overline M}_\alpha 
\Gamma_{ij} M_\alpha~, \label{kmono1} \\
&& \sum_i \Gamma^i D_i M_\alpha = 0~, \hskip10mm \alpha = 1, 2, \cdots, k~.
\label{kmono2}
\eeqa
Using the Weitzenb\"ock formula for each component $M_\alpha$,
we obtain
\beqa
& & \int_X d^4 x \sqrt{g} \left( \frac{1}{2} \left| F^+ + \frac{i}{2} 
\sum_{\alpha=1}^{k} {\overline M}_\alpha \Gamma M_\alpha \right|^2
+ \sum_{\alpha=1}^{k} \left| \Gamma \cdot D M_\alpha \right|^2
\right) \CR
& & =\int_X d^4 x \sqrt{g} \left( \frac{1}{2} | F^+ |^2 +  \sum_{\alpha=1}^{k} 
\left| DM_\alpha \right|^2 + \frac{1}{4} R ( \sum_{\alpha=1}^{k} 
{\overline M}_\alpha M_\alpha ) + \frac{1}{2} 
( \sum_{\alpha=1}^{k} {\overline M}_\alpha M_\alpha )^2 \right),
\eeqa
where $R$ is the scalar curvature.
Since
\beq
 \int_X d^4 x \sqrt{g} 
\left( \sum_{\alpha=1}^{k} {\overline M}_\alpha M_\alpha 
+ \frac{1}{4} R \right)^2 \geq 0~,
\eeq
we have the following bound;
\beq
\int_X d^4 x \sqrt{g} \left( \frac{1}{2} | F^+ |^2 +  \sum_{\alpha=1}^{k} 
| DM_\alpha |^2 \right)  \leq \frac{1}{32} \int_X d^4 x \sqrt{g} R^2~,
\eeq
if the generalized monopole equations (\ref{kmono1}) and (\ref{kmono2}) are
satisfied.
Thus we obtain a bound 
\beq
I_+ = \int_X d^4 x \sqrt{g}  | F^+ |^2  \leq \frac{1}{16} \int_X d^4 x
\sqrt{g} R^2~.
\label{bound1}
\eeq
Furthermore, note that
\beq
c_1(L^2)^2 = \frac{1}{(2\pi)^2} \int_X d^4 x \sqrt{g}  
\Big( | F^+ |^2 - | F^- |^2 \Big)~.
\eeq
Therefore, in order for the formal dimensions $d_\lm$ of the moduli space 
to be non-negative, we must have
\beq
\frac{1}{(4\pi)^2} \int_X d^4 x \sqrt{g}  
\Big( | F^+ |^2 - | F^- |^2 \Big)  \geq \frac{\sigma}{4} + \frac{1}{2k}
(\sigma + \chi)~.
\eeq
Hence we have another bound 
\beq
I_- = \int_X d^4 x \sqrt{g}  | F^- |^2  \leq \frac{1}{16} \int_X d^4 x
\sqrt{g} R^2
- 4 \pi^2 (\sigma+ \frac{2}{k}(\sigma + \chi))~.
\label{bound2}
\eeq
Since both $I_+$ and $I_-$ are bounded, the set of $\lm$ with 
non-vanishing $SW(\lm)$ is in a compact subset of $H^2(X, \bf{R})$.
Hence, there are only finite $Spin^c$ structures that have
non-vanishing contribution to the Donaldson-Witten function.

For example, we  can take $R=0$ for $K3$ surface which is hyper K\"ahler.
Substituting $\chi =24$, $\sigma =-16$, we obtain
\beq
I_+ = 0~, \quad I_- \leq 64 \pi^2 \Big( 1 - \frac{1}{k} \Big)~.
\eeq
For $k=1$ we find a well-known fact that on a $K3$ surface
only a trivial class $\lm=0$ has non-vanishing contributions,
which is the basic class in the sense of Seiberg-Witten.
On the other hand for $k=2,4$, this is no longer true.
By taking into account that the minimum of $J = \int_X F \wedge F$ is
 $8(2\pi)^2$ \cite{W3}, (note also that $\Gamma = H_2 (K3, \bf{Z})$ is
an even lattice, since $K3$ is spin),
we see that non-vanishing contributions come from either a trivial
class $\lm=0$ or classes with $\lm^2 = -2$. 
Since a $K3$ surface has $b_2^- = 19$, we conclude that there are
$1+ 2 \cdot19$ possible classes.

\section{$K3$ surface}

\renewcommand{\theequation}{4.\arabic{equation}}\setcounter{equation}{0}

As the simplest example let us calculate $Z_{DW}$ for a $K3$ surface for which
$\chi =24$, $\sigma =-16$ and hence $d_\lambda =k\lambda^2+4(k-1)$. Since
$b_2^+=3$ the $u$-plane integral vanishes. Let us assign the $R$ charge 4
to $p$ and 2 to $S$, then the degree $s$ Donaldson polynomial contains 
terms $S^np^t$ with $2n+4p=s$ where $s$ is the dimension of the instanton
moduli space. In the case of massless $N=2$ $SU(2)$ QCD on a $K3$ surface,
we have 
\beq
s=2(4-N_f)\ell -12+8N_f,
\eeq
where $\ell$ is the instanton number. For $N_f \geq 1$, it is important to
recall here that only the instantons with even instanton number contribute
due to the anomalous ${\bf Z}_2$ symmetry \cite{SeWi2}.
To evaluate $Z_{SW}$ one has to choose a
good local coordinate near each singularity on the $u$-plane. A good
coordinate is given by $ga_D+qa$ where $g$ and $q$ are the magnetic and
electric charges of the BPS dyonic state which becomes massless at the
singularity. In what follows we shall set $\La_{N_f}=1$ for simplicity.

We start with the well-known $N_f=0$ theory. Only $\lambda^2=0$ contributes
to $Z_{SW}$, and the famous expression results in \cite{OG}\cite{W4}
\beq
Z_{DW}={c_0 \over 2}
\left(e^{2p+\half S^2}-e^{-2p-\half S^2} \right)
=c_0 \sinh \left(2p+\half S^2\right),
\eeq
where $c_0$ is an overall constant and the two terms are due to a 
${\bf Z}_2$ pair of singularities 
at $u=\pm 1$. Note that the terms of the $R$ charge congruent to 4 modulo 8 
exist in accordance with the selection rule with $s=8\ell -12$.
$Z_{DW}$ satisfies the so-called simple type condition
\beq
\left( {\pa^2 \over \pa p^2}-4\right) Z_{DW}=0.
\eeq

In the $N_f=1$ theory there appear singularities in a ${\bf Z}_3$ symmetric
manner at $u=-2^{1/3}\,3\, \omega^j/8$ with $j=0,\, 1, \, 2$ and 
$\omega =e^{2\pi i/3}$. 
At each singularity a single massless monopole (or dyon)
comes out as a matter hypermultiplet. The property of the resulting $k=1$ 
monopole equation is essentially the same as the $N_f=0$ case. 
Thus only the class $\lambda^2=0$ contributes, leading to the result \cite{LM}
\beq
Z_{DW}=c_1 \sum_{j=0}^2 \omega^j 
e^{\alpha \omega^j \left(2p+{2 \over 3}S^2\right)},\hskip10mm
\alpha =-2^{1/3} {3\over 8},
\label{nf3dw}\eeq
where surviving terms on the RHS have the $R$ charge 8 modulo 12. This is 
in agreement with the selection rule obeyed by the even instanton
contributions. As pointed out in \cite{LM}, $Z_{DW}$ is subject to the
equation
\beq
\left( {\pa^3 \over \pa p^3}+{27 \over 32} \right) Z_{DW}=0.
\eeq
Note that (\ref{nf3dw}) can be reproduced by taking 
the massless limit of $Z_{DW}$ for the massive
$N_f=1$ theory \cite{MoWi}, since the number of massless particles at
singularities does not change in the limit. On the other hand, this limit
becomes singular in the $N_f \geq 2$ theory to which we next turn.

In the $N_f=2$ theory the ${\bf Z}_2$ symmetry is acting on the $u$-plane. 
At the singularity $u=1/8$, we have massless monopoles in 
${\bf 2_s}$ of $SO(4)$ and
a good local coordinate is $a_D$ or $q_D=e^{2\pi i\tau_D}$ with 
$\tau_D=-1/\tau$. The other singularity is located
at $u=-1/8$ where dyons in ${\bf 2_c}$ of $SO(4)$ with the charge 
$(g,q)=(1, -1)$ are massless. A good coordinate is 
thus $a_D-a$, and the corresponding modular expressions
are obtained by letting $\tau \rightarrow \tilde\tau =\tau -1 \rightarrow
\tilde\tau_D =-1/\tilde\tau$ in (\ref{modular}). The relevant $q$-expansion
is then performed around $\tilde q_D=0$ where 
$\tilde q_D=e^{2\pi i \tilde\tau_D}$. Furthermore, 
in calculating $Z_{SW}$ we have to take the contributions of $\lambda^2 =-2$ 
($d_\lambda =0$) in addition to $\lambda^2 =0$ ($d_\lambda =4$) 
as discussed in the preceding section. We now obtain from
(\ref{ours}) that
\beqa
&& Z_{DW}=
c_2\, \Bigl[ \left( 16p^2+8pS^2+S^4-88p-24S^2+228 \right)\, e^{p/4+S^2/8}  \no
&& \hskip15mm 
-\left( 16p^2+8pS^2+S^4+88p+24S^2+228 \right)\, e^{-p/4-S^2/8} \Bigr] \no
&& \hskip15mm 
+{c_2}' \sum_{\lambda^2 =-2} SW(\lm)
\left( e^{p/4+S^2/8+\sqrt{2}\, (S,\lambda)}
-e^{-p/4-S^2/8-i\sqrt{2}\, (S,\lambda)} \right),
\eeqa
where $c_2,\, {c_2}'$ are constants. When we sum over $\lm$ the symmetry 
property \cite{W2}\cite{Mor}
\beq
SW(-\lm)=(-1)^\delta SW(\lm)
\eeq
should be taken into account. It is evident that
\beq
\left( {\pa^2\over \pa p^2}-{1\over 16}\right)^3 Z_{DW}=0.
\label{DWcond}
\eeq
Thus, when the massless $N_f=2$ $SU(2)$ theory is considered on $K3$ surfaces,
$Z_{DW}$ obeys the condition which is reminiscent of the one for manifolds of
generalized simple type in the sense of \cite{KM}. 

Finally, in the $N_f=3$ theory, there is no symmetry acting on the $u$-plane. 
Massless monopoles in ${\bf 4}$ of $SO(6)$ appear 
at the singularity $u=0$ around which we take $a_D$
or $q_D$ as a good coordinate. An additional singularity associated with an
$SO(6)$ singlet massless dyon with the charge $(2,1)$ exists 
at $u={1\over 256}$.
A good coordinate near this singularity is taken to be $2a_D+a$. 
Correspondingly we implement transformations $\tau \rightarrow 
\tilde\tau =\tau +1/2 \rightarrow \tilde\tau_D=-1/\tilde\tau$ in 
(\ref{modular}) so that we can make the power series expansions at
$\tilde q_D=0$. For instance we find
\beqa
u &=& {1\over 256} {\th_3^4 \left({\tilde\tau_D \over 2} \right) \over
\left( \th_3^2-\th_2^2 \right)^2 (\tilde\tau_D)} \no
&=&{1\over 256} \left( 1+16 \tilde q_D^{1/4}+128 \tilde q_D^{1/2}+
744 \tilde q_D^{3/4}+3072 \tilde q_D +\cdots \right).
\eeqa
The singularities at $u=0$ and $u={1\over 256}$ respectively give rise to 
$Z_{SW}^{(1)}$ and $Z_{SW}^{(2)}$, thereby the SW contribution is obtained as
$Z_{SW}=Z_{SW}^{(1)}+Z_{SW}^{(2)}$.
As we have shown, $Z_{SW}^{(1)}$ consists of the contributions of 
$\lambda^2=0$ ($d_\lambda =12$) and $\lambda^2=-2$ ($d_\lambda =4$), while
only $\lambda^2=0$ contributes to $Z_{SW}^{(2)}$ since the monopole equation
is multiplicity free ({\it i.e.} $k=1$). Our result reads
\beqa
 Z_{SW}^{(1)} &=& c_3 \Biggl(
{1\over 2949120}p^6+{1\over 188743680}S^{12}+{1\over 3145728}p^2 S^8
+{1\over 15728640}p S^{10}+{1\over 983040}p^5 S^2  \no
&& +{1\over 786432}p^4 S^4+{1\over 1179648}p^3 S^6+78648p+72772S^2+10879744
+{521 \over 2}p^2 \no
&&+{2101 \over 4}p S^2+241 S^4+{213 \over 128}p^2 S^2+{441 \over 256}p S^4
+{31 \over 64}p^3+{809 \over 1536}S^6+{3\over 512}p^3 S^2 \no
&& +{3\over 512}p^2 S^4+{23 \over 6144}p S^6+{7 \over 1536}p^4
+{7\over 8192}S^8-{29\over 98304}p^4 S^2-{7\over 32768}p^3 S^4 \no
&&-{13 \over 196608}p^2 S^6-{5\over 786432}p S^8-{37 \over 245760}p^5
+{1\over 2621440}S^{10} \Biggr) \no
&&+c_3' \sum_{\lambda^2=-2} SW(\lm)
\Bigl( 1456p-14208 \sqrt{2}\, i(S,\lambda)+792S^2+4p^2
-64\sqrt{2}\, ip (S,\lambda) \no
&& +4p S^2-512 (S,\lambda)^2-32\sqrt{2}\, i (S,\lambda)S^2+S^4+220672 \Bigr) \,
e^{i \sqrt{2}\, (S,\lambda)/4},
\no
 Z_{SW}^{(2)} &=&
c_3''\, \Bigl( 64p^6+S^{12}+60p^2S^8+12p S^{10}+192p^5 S^2+240 p^4 S^4
+160 p^3 S^6 \no
&& -15850442588160p-15061242347520 S^2+2218796211240960 \no 
&&+50520391680p^2 +108338872320p S^2+50787778560S^4-332267520p^2 S^2 \no
&&-366673920pS^4-74711040p^3-113541120S^6+655360p^3 S^2+1105920p^2S^4 \no
&&+819200pS^6+409600p^4+189440S^8+48640p^4S^2+33280p^3S^4+8960p^2S^6 \no
&&+320pS^8+25600p^5-160S^{10} \Bigr) \, e^{(p+S^2)/128}.
\eeqa
Hence,
\beq
{\pa^7 \over \pa p^7} \left( {\pa \over \pa p}-{1\over 128}\right)^7 
Z_{DW}=0.
\eeq

\section{Discussions}

\renewcommand{\theequation}{5.\arabic{equation}}\setcounter{equation}{0}

The Donaldson-Witten functions we have obtained for a $K3$ surface
involve $SW(\lm)$ for $\lm^2=-2$ and constants $c_i,\, c''_3$ which are fixed 
by the values of $SW(\lm)$ for $\lm^2=0$.
To establish mathematically more rigorous basis, we have to prove
the compactness of the moduli space ${\cal M}_\lm$ so that the integral
$SW(\lm)$ in (\ref{SWinv}) is well-defined.\footnote{We thank H. Nakajima for 
pointing out this issue to us.} This point is more subtle than the case
of the bounds for $I_\pm$ we have shown in section 3.3. 
What remains is to show a similar bound for the monopole fields $M_\alpha$. 
In the case of the Seiberg-Witten monopole
equation with a single massless monopole, such a bound is obtained
rather easily \cite{Mor}. 
(In fact this is one of advantages of the abelian nature
of Seiberg-Witten invariants.) 
However, the genelarized monopole equation with multi-component
massless monopoles does not seem to allow a straightforward
generalization of the argument in \cite{Mor}. At the moment
we have assumed the compactness of the moduli space and leave the proof as 
an open problem.

When there are multiple massless monopole fields,
we had to pick up higher order terms in the $q$-series expansion
near the singularities as the contribution from the trivial class $\lm$.
This is also expected when a four-manifold $X$ does not satisfy the simple
type condition. Thus the expansions of elliptic modular functions
near the strong-coupling singularities we have given in the paper
may be useful for computing the Donaldson invariants for 
a manifold of generalized simple type, though at present it is only
a hypothetical case when $b_2^+ > 1$.

Finally we point out 
an interesting issue related to twisted $N=2$ superconformal
field theory in four dimensions. In \cite{MM2} Mari\~no and Moore initiated the
study of four-dimensional topological conformal theories by twisting the $N=2$
superconformal theory realized at the Argyres-Douglas point in the $SU(3)$
Yang-Mills theory \cite{AD} and its generalization to $SU(N)$ \cite{EHIY}. 
The $N=2$ superconformal fixed point which is in the same 
universality class with
the Argyres-Douglas point is known to exist in the massive $N=2$ $SU(2)$
QCD with $N_f=1$ \cite{APSW}. Moreover, analogous non-trivial fixed points are
found in $N=2$ $SU(2)$ QCD with $N_f>1$ massive flavors \cite{APSW}.
These fixed points are also obtained in $N=2$ pure Yang-Mills 
theories \cite{EHIY}.
We then expect that, though microscopic theories are distinct, the $N=2$
fixed points in the same universality class should yield the identical
topological conformal theory after being twisted. Thus, it will be worth
working out this relationship explicitly  using massive $N=2$ $SU(2)$ QCD and
$N=2$ pure Yang-Mills theories. This may shed new light not only on topological
conformal theory, but on four-dimensional $N=2$ superconformal theory whose
precise dynamics still needs to be clarified.

\noindent
{\Large\bf Acknowledgements}

\vskip3mm\noindent
We would like to thank H. Nakajima for helpful communications.
This work was supported in part by Grant-in-Aid for Scientific
Research from the Ministry of Education, Science and Culture
(Nos. 10640081, 09640335).

\newpage

\noindent
{\Large\bf Appendix. Periods and modular functions}

\renewcommand{\theequation}{A.\arabic{equation}}\setcounter{equation}{0}

\vskip3mm\noindent
In order to evaluate $Z_{SW}$ we need to express various quantities in terms
of modular forms. For this we first identify the roots of the cubic in
(\ref{curve}) with $e_\nu$ by examining the large-$u$ asymptotic behavior.
Then, after some algebra, we find necessary formulas. Including the $N_f=0$
case \cite{MoWi} to make our paper self-contained, 
let us summarize our results;
\beqa
{N_f=0 \over } \hskip5mm
&& {u\over \La_0^2}={\th_2^4+\th_3^4 \over 2(\th_2\th_3)^2},
\hskip10mm {2\omega_1\La_0 \over \pi}=2\sqrt{2} \th_2\th_3, \no
&& P_0(u)=\La_0^4 {\th_4^8\over (\th_2\th_3)^4}, \hskip10mm
{d\tau \over du}={4i \over \pi \La_0^2} {(\th_2\th_3)^2 \over \th_4^8}, \no
{N_f=1 \over } \hskip5mm
&& {u\over \La_1^2}={2\sqrt{2f} \over A}, 
\hskip10mm {2\omega_1\La_1 \over \pi}= \sqrt{A},
\hskip10mm A^3={2^{11} \over 3^3}\left( 2h-f\sqrt{2f} \right), \no
&& P_1(u)=-2^{17} \La_1^4 {(\th_2\th_3\th_4)^8\over A^5 \sqrt{2f}}, \hskip10mm
{d\tau \over du}=-{3i \over 2^{14}\pi \La_1^2}
{\sqrt{2f} A^4 \over (\th_2\th_3\th_4)^8}, \no
{N_f=2 \over } \hskip5mm
&& {u\over \La_2^2}={\th_3^4+\th_4^4 \over 8\th_2^4},
\hskip10mm {2\omega_1\La_2 \over \pi}=4\th_2^2, \no
&& P_2(u)=\La_2^4 {(\th_3\th_4)^4 \over 4\th_2^8}, \hskip10mm
{d\tau \over du}={4i \over \pi \La_2^2}{\th_2^4 \over (\th_3\th_4)^4}, \no
{N_f=3 \over } \hskip5mm
&& {u\over \La_3^2}=-{(\th_3\th_4)^2 \over 64 (\th_3^2-\th_4^2)^2}, 
\hskip10mm {2\omega_1\La_3 \over \pi}=16i (\th_3^2-\th_4^2), \no
&& P_3(u)={\La_3^4 \over 64^2} {(\th_3\th_4)^2 \over (\th_3^2-\th_4^2)^4}
(\th_3^2+\th_4^2)^2, \hskip10mm
{d\tau \over du}=-{128 i \over \pi \La_3^2}
{(\th_3^2-\th_4^2)^2 \over (\th_3\th_4)^2 (\th_3^2+\th_4^2)^2}, \no
&&
\label{modular}
\eeqa
where modular functions $f$ and $h$ are defined in (\ref{fandh}).

As a by-product the $\beta$ function defined by
\beq
\beta (\tau)=\La {\pa \tau \over \pa \La}\biggr|_{u\; {\rm fixed}}
=-2u {\pa \tau \over \pa u}\biggr|_{\La \; {\rm fixed}}
\eeq
is obtained including full instanton corrections. We find
\beqa
{N_f=0 \over } \hskip5mm
&& \beta (\tau)={4 \over \pi i}{\th_2^4+\th_3^4 \over \th_4^8}
={4 \over \pi i}\left( 1+40q^{1/2}+552q+4896q^{3/2}+\cdots \right), \no
{N_f=1 \over } \hskip5mm
&& \beta (\tau)={i \over 9\pi} {f \left( 2h-f\sqrt{2f} \right) 
\over (\th_2\th_3\th_4)^8}
={3 \over \pi i}\left( 1+312q+20520q^2+497760q^3+ \cdots \right), \no
{N_f=2 \over } \hskip5mm
&& \beta (\tau)={1 \over \pi i} {\th_3^4+\th_4^4 \over (\th_3\th_4)^4}
={2 \over \pi i}\left( 1+40q+552q^2+4896q^3+\cdots \right), \no
{N_f=3 \over } \hskip5mm
&& \beta (\tau)={4 \over \pi i}{1 \over (\th_3^2+\th_4^2)^2}
={1\over \pi i}\left( 1-8q+40q^2-160q^3+\cdots \right).
\eeqa
Note that the leading term $(4-N_f)/\pi i$ is the one-loop $\beta$ function.
Note also that only integral powers of $q$ appear for $N_f \ge 1$. This
is due to the fact that only even instantons contribute in $N_f \ge 1$
theories \cite{SeWi2}. The results agree with \cite{MiNe}\cite{beta} 
for $N_f=0$, and with \cite{MiNe} for $N_f=1,\, 2$. 

The SW periods (\ref{period}) are also expressed in terms of modular functions.
We use (\ref{modular}) and the relations
\beq
\zeta (\omega_1)={\pi^2 \over 12 \omega_1}E_2(\tau), \hskip10mm
\zeta (\omega_3)=-{\pi^2 \over 12 \omega_1}\tau_D E_2(\tau_D),
\eeq
where $\tau_D=-1/\tau$ and the normalized Eisenstein series
\beq
E_2(\tau)=1-24 \sum_{n=1}^\infty {nq^n \over 1-q^n}
\eeq
of weight 2 is transformed as
\beq
E_2(\tau)={6 \tau_D \over \pi i}+\tau_D^2 E_2(\tau_D)
\eeq
under $\tau \rightarrow \tau_D$. We then obtain from (\ref{period}) that
\beqa
{N_f=0 \over } \hskip5mm
&& {a \over \La_0}={1\over 6} {2E_2+\th_2^4+\th_3^4 \over \th_2\th_3},
\hskip10mm 
{a_D \over \La_0}=-{i \over 6} 
\left( {2E_2-\th_3^4-\th_4^4 \over \th_3\th_4} \right) (q_D), \no
{N_f=1 \over } \hskip5mm
&& {a \over \La_1}={1\over 2\sqrt{2}}{2E_2+\sqrt{2f} \over \sqrt{A}},
\hskip10mm
{a_D \over \La_1}={1 \over 2\sqrt{2}}
\left( {2E_2-\sqrt{2f}\over \sqrt{A_D}}\right) (q_D),         \no
&& \hskip30mm A_D^3={2^{11}\over 3^3} 
\left( 2h+f\sqrt{2f}\right) (q_D), \no
{N_f=2 \over } \hskip5mm
&& {a \over \La_2}={1\over 6\sqrt{2}} {E_2+\th_3^4+\th_4^4 \over \th_2^2},
\hskip10mm
{a_D \over \La_2}={1\over 6\sqrt{2}i}
\left( {E_2-\th_3^4-\th_2^4 \over \th_4^2} \right) (q_D), \no
{N_f=3 \over } \hskip5mm
&& {a \over \La_3}={1\over 48\sqrt{2}i}
{E_2+3(\th_3\th_4)^2+\th_3^4+\th_4^4 \over \th_3^2-\th_4^2}, \no
&& {a_D \over \La_3}=-{1 \over 48\sqrt{2}}
\left( {E_2-3(\th_3\th_2)^2-\th_3^4
-\th_2^4 \over \th_3^2-\th_2^2}\right) (q_D),
\eeqa
where $a$ is presented as a function of $q$ while $a_D$ as a function of
$q_D=e^{2\pi i \tau_D}$.

Finally the contact term is given by \cite{MM1}
\beq
T(u)=-{1\over 24}E_2(\tau)\left( {du \over da}\right)^2
+{1\over 3}\left( u+{\La_3^2 \over 64}\delta_{N_f,3}\right).
\eeq
Expressing this in terms of modular functions is immediate with the use of
(\ref{modular}).

\newpage
 

\end{document}